\documentclass[final,3p,times]{elsarticle}

\usepackage{amssymb,amsmath,amsthm, amsfonts, amstext}
\usepackage{footnote}
\usepackage{rotating}
\usepackage{setspace}
\usepackage{hhline}
\usepackage{hyperref}
\usepackage[normalem]{ulem}
\usepackage{multirow}
\usepackage{booktabs}
\usepackage{graphicx}
\usepackage{chngcntr}
\usepackage{color}
\usepackage{soul}
\usepackage{url} 
\usepackage{lineno}

\counterwithout{equation}{section}

\journal{Elgar}

\nonstopmode

\begin{document}

\begin{frontmatter}

\title{Cryptocurrencies and the Future of Money  }

\author[a1]{Matheus R. Grasselli}
\author[a2]{Alexander Lipton}

\address[a1]{McMaster University, Hamilton, Canada}
\address[a2]{MIT, Cambridge, USA}

\begin{abstract}
We review different classes of cryptocurrencies with emphasis on their economic properties. Pure-asset coins such as Bitcoin, Ethereum and Ripple are characterized by not being a liability of any economic agent and most resemble commodities such as gold. Central bank digital currencies, at the other end of the economic spectrum, are liabilities of a Central Bank and most resemble cash. In between, there exist a range of so-called stable coins, with varying degrees of economic complexity. We use balance sheet operations to highlight the properties of each class of cryptocurrency and their potential uses. In addition, we propose the basic structure for a macroeconomic model incorporating all the different types of cryptocurrencies under consideration.   
\end{abstract}


\end{frontmatter}



\section{Introduction}
\label{introduction}

Cryptocurrencies, and the Distributed Ledger Technology (DLT) on which they are based, have captured the attention and imagination of investors, academics, politicians and the general public alike. Touted by technology enthusiasts as the future of money, cryptocurrencies promised advantages include a secure and completely decentralized payment system, a drastically simplified and more stable financial sector, and a more inclusive, transparent and democratic economy. Reality has been more nuanced and marked by high volatility in the price of well-known cryptocurrencies such as Bitcoin, an uncomfortable degree of opacity and centralization, and inefficient implementations, in particular with respect to energy consumption. The allure of using the latest developments in computer science and cryptography to solve time-honoured economic problem is, however, too great to ignore. Accordingly, despite the glaring difficulties just mentioned, novel applications of DLT continue to be proposed, including revolutionizing the way artwork and other intellectual property are traded through the use of non-fungible tokens. 

The full breath of applications of cryptocurrencies, blockchains and DLT, as well as the technical background needed to understand them, are of course beyond the scope of any single article, and we recommend \cite{LiptonTreccani2021} as an entry point to this fascinating world. In this paper, we restrict ourselves to a discussion of the strictly monetary applications of DLT. Specifically, we consider: pure-asset, or commodity-like, cryptocurrencies such as Bitcoins in Section \ref{bitcoin_section}; central bank digital currencies (CBDC) in Section \ref{CBDC_section}; and the more recent class of so-called stable coins in Section \ref{stable_section}. For each of these types of cryptocurrencies, we provide a definition, a brief description of the technical aspects underlying each of them, and some prominent representative cases. In each case, we explore their economic characteristics through an extensive use of balance sheet operations in the manner of \cite{Lavoie2003}. In Section \ref{SFC}, we describe the type of macroeconomic framework that best aggregates these balance sheet operations in a coherent way, namely a stock-flow consistent model incorporating the different types of cryptocurrencies.

\section{Pure-asset coins}
\label{bitcoin_section}

There are many overlapping and non-equivalent ways to classify cryptocurrencies, such as their degree of centralization and anonymity, or the type of algorithm used for validation and consensus building \cite{TascaTessone2019}. As mentioned in the previous section, because we are primarily interested in the economic aspects of cryptocurrencies, we adopt a taxonomy based on their balance-sheet status. That is to say, we focus on what type of assets and liabilities they represent and for which economic agents.   

Economically speaking, the simplest type of cryptocurrency consists of what we call pure-asset coins. This is also the original and, at present, most widespread class of cryptocurrencies and includes not only the Big Three well-known examples of Bitcoin (BTC), Etherium (ETH) and Ripple (XRP), but also many precursors such as Ecash, Digicash, bit gold, and b-money, as well as subsequent variations such as Litecoin and countless replicas, or {\em alt-coins}, amongst them Dogecoin. 

From an economic point of view, the defining feature of this type of cryptocurrencies is that they are not a liability of any specific agent. That is to say, they figure on the balance sheet of the agents holding them as an asset, while they do not figure as a liability in the balance sheet of any other economic agent. In this respect, cryptocurrencies of this type are similar to physical commodities such as gold or silver, and should {\em not} be viewed as a financial asset such as a bond or a stock. Much like physical commodities, the simplest way to acquire pure-asset cryptocurrencies is by exchanging them for other assets (say purchasing gold using bank deposits) or by providing goods and services (say by being paid in gold for delivering a really good musical performance). Still analogous to physical commodities, another way to acquire pure-asset cryptocurrencies is through {\em mining}, that is to say, by executing a specific task that is associated with the creation of new coins---the term used for a unit of cryptocurrency---that did not exist before. Accordingly, the difficulty and related cost (e.g. in terms of required time and resources) associated with mining affect the available supply of such coins, and consequently their price.   

Once acquired, a pure-asset coin can then be exchanged by other assets (say by selling them to someone in exchange for some amount of bank deposits) or for other goods and services (say using them to pay a dealer for some artwork). In all cases, because they are not a liability for any economic agent, nobody has the obligation to covert a pure-asset coin into anything else. Moreover, unlike physical commodities, pure-asset coins do not have any intrinsic value for non-monetary purposes, for example in the way that gold can be used in dentistry. As a result, the value of such coins depend exclusively on the willingness of other agents accepting them in exchange of other assets, goods, or services. 

Take Bitcoin as an illustrative example. The decentralized, permissionless Bitcoin protocol, described for the first
time in the seminal paper by Satoshi Nakamoto \cite{Nakamoto2008}, launched the blockchain
revolution. Nakamoto
articulated his objective as follows:

\textquotedblleft I've been working on a new electronic cash system that's
fully peer-to-peer, with no trusted third party --- The main properties:
Double-spending is prevented with a peer-to-peer network. No mint or other
trusted parties. Participants can be anonymous. New coins are made from
Hashcash style proof-of-work. The proof-of-work for new coin generation also
powers the network to prevent double-spending.\textquotedblright 

With some simplifications made for the sake of clarity, the protocol can be
described as follows (see \cite[Chapter 5]{LiptonTreccani2021} for more details). The protocol uses a native token called Bitcoin or
BTC. The main objective of the protocol is to make sure that ownership of
BTC is internally consistent. The protocol says nothing about the value of
BTCs in USD or other fiat currencies. This value is established by supply
and demand considerations on (mostly centralized) exchanges, operating
entirely outside the protocol itself. At its peak, the value of one BTC was
63,500 USD. 

To participate in the protocol, one must create a public-secret key pair,
which one can do without asking anyone's permission; hence, the protocol is
permissionless. The public key is a point $P$ on the Koblitz elliptic curve,
secp256k1; the secret key is a number $k$, such that $P=k \times G$, where $G$ is the
so-called base point. As their names suggest, public keys are known to all
the participants. In contrast, secret keys must be protected at all costs
since losing a secret key is tantamount to losing BTCs associated with the
corresponding public key. Collections of key pairs are called Bitcoin wallets. As of
this writing, there are about 200 million Bitcoin wallets. Copies of the
Bitcoin ledger are controlled by the so-called full nodes, which use what is termed proof-of-work (PoW) consensus protocol (as described below) to ensure that
all these copies are mutually consistent. Presently, there are about
thirteen thousand full nodes. In addition, there are Bitcoin miners whose
role (as explained below) is to maintain the integrity of the Bitcoin ledger. 

Not surprisingly, the concept behind the Bitcoin protocol is purely
mathematical. Bitcoin grows by induction, starting with the so-called
genesis block. The principal Bitcoin operation transfers BTCs from one
wallet (address) to the following wallet (address). These transfers are
cryptographically secured and have to be signed by the secret key of the
donor address. Thus, Bitcoin transactions are organized as chains, with the
output of one transaction becoming the input of the next one. These chains
can be traced back to the so-called Coinbase transactions (as explained below),
or even further to the genesis block.

Of course, the entire Bitcoin ledger exists only online. Hence, if no
additional measures are taken, it would naturally suffer from the so-called
double-spending. For example, the owner of wallet A can send her BTCs
simultaneously to two (or even many) wallets, say B and C. This possibility
does not occur when payment is made in cash since A can physically give her
money either to B or C, but not both. However, it can happen if A writes two
checks but has funds to cover only one; this illegal practice is known as
check kiting. It also does not occur when electronic transfers are performed
by authorized third parties, such as banks. Since Bitcoin is entirely
self-contained and does not have exogenous intermediaries, the protocol has
to control the validity of transfers endogenously. This control is exercised
by miners, who maintain the integrity of the ledger. Here is how they
operate.

\begin{figure}[h!]
\centering
   \includegraphics[trim=0 3cm 0 3cm,width=\textwidth]{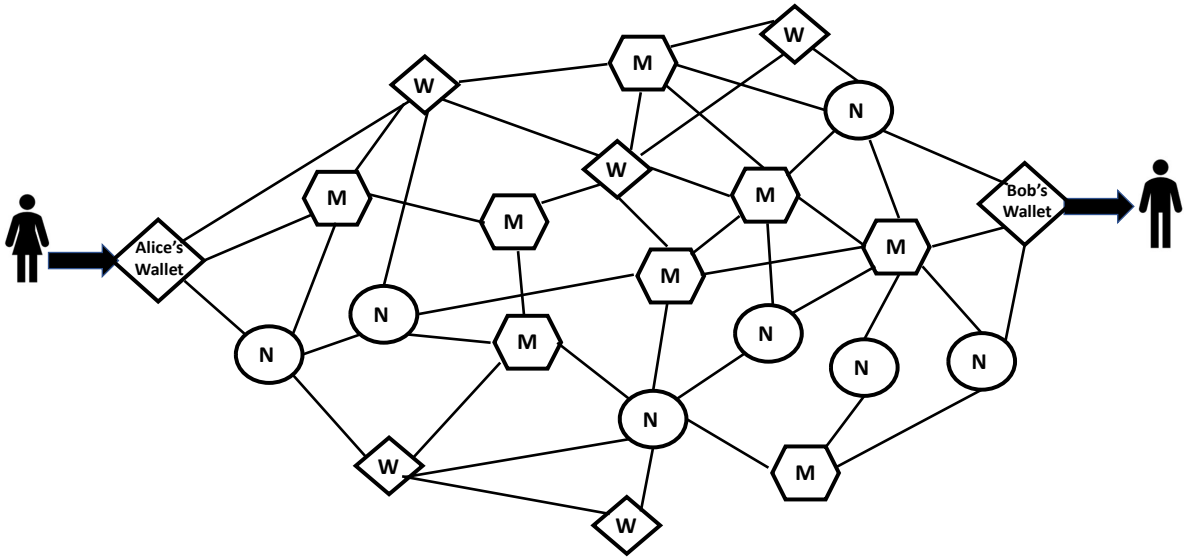}
 \caption{How Alice pays Bob via the Bitcoin protocol. N denotes full nodes, M denotes miners, and W denotes wallets.}
 \label{ecosystem_figure}
\end{figure}

\begin{figure}[h!]
\centering
   \includegraphics[trim = 0 3cm 0 3cm,width=\textwidth]{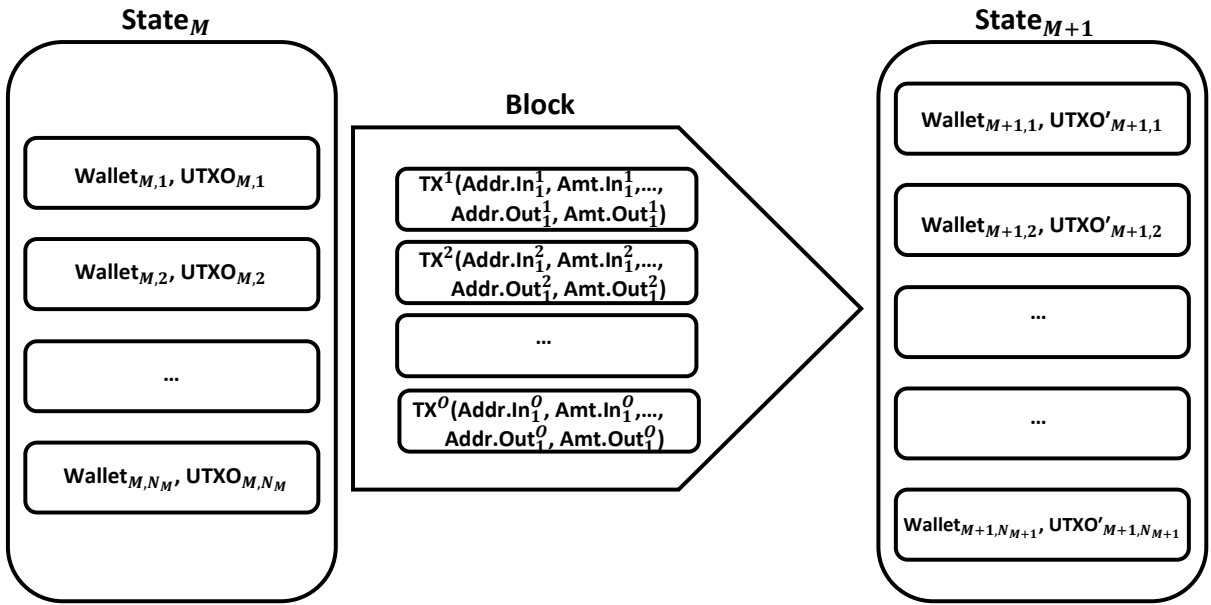}
 \caption{The Bitcoin ledger is a giant Markov chain. Its state changes when a new block is added to it.}
 \label{transition_figure}
\end{figure}

When the owner of wallet A (traditionally called Alice) wishes to transfer
her BTCs to the owner of wallet B (traditionally called Bob), she broadcasts
her intention to the network of full nodes and miners. Her transaction goes
into the memory pool. Miners extract transactions from the pool and form
them into blocks. On average, a block contains two thousand transactions.
First, miners check that all proposed transactions in their blocks are
correctly signed, have enough funds in the donor wallet, etc. It is
important to emphasize that miners receive transactions at different times
so that timestamps cannot order transactions properly. Hence, blocks created
by different miners are not identical. Once the block's content is verified,
miners compete to validate their respective blocks as soon as possible.
Participating in block validation requires enormous resources, including
hardware, software, personnel, and, above all, electricity. This is why the
corresponding validation is said to be based on PoW. Figure \ref{ecosystem_figure} illustrates the above description graphically. Figure \ref{transition_figure} shows how the Bitcoin ledger evolves when a new block is added to it.

Moreover, since the validation game relies on the winner-takes-all
principle, all participants spend these resources. As a result, the
electricity wastage by the protocol is genuinely gargantuan. Bitcoin
currently consumes around 110 Terawatt Hours per year\footnote{See \url{https://cbeci.org}}, equal to the annual
energy consumption of Sweden, or 0.55\% of the world's total. Bitcoin has a
truly clever mechanism for controlling the complexity required to validate a
block. When the number of miners increases or decreases, so does the
complexity of the validation. As a result, on average, it always takes 10
minutes to validate a block. Of course, miners spend all the resources
because the successful miner is rewarded with the block reward, paid by the
protocol via the Coinbase transaction mentioned earlier. Rewards are halved
every four years (on average). Initially, they were 50 BTC per block.
Currently, rewards are 6.25 BTC per block. Due to the periodic halving of
the Coinbase rewards, the asymptotic value of the total supply is 21 million
BTCs. The PoW protocol allows Bitcoin ledgers observed by
different full nodes to agree, at least for blocks sufficiently deep with
the blockchain. If ledgers start to disagree at the top (for the most recent
blocks), the longest valid chain wins. Thus, a participant in the Bitcoin protocol can receive BTC into her address
in two ways, either by mining or from an address, which acquired BTC
previously.

Tables \ref{bitcoin1} to \ref{bitcoin3} illustrate the transactions mentioned above. In Table \ref{bitcoin1} we see that the act of mining by Alice leads to an increase of \$10,000 in the Bitcoin asset class, without either an increase in liabilities or a decrease in any other asset, except for the deposit account, which is used to pay for the operational costs of mining (e.g electricity) of \$9,000. As a result, Alice's net worth increases by the amount of Bitcoin that is mined, net of costs, or \$1,000 in this example. In all the examples that follow, we highlight in bold the affected balance sheet entries after a given transaction. Observe that, from the assets listed in this example, Bitcoin most resembles a house and a car, in the sense that they are not liabilities for any other economic agent, wheres bank deposits and savings are liabilities for Alice's bank. 

\begin{table}[h]
\centering
\begin{tabular}{ lr | lr } 
\multicolumn4c{\bf Alice's Balance Sheet (before)} \\
\toprule
\multicolumn2c{Assets (\$)} & \multicolumn2c{Liabilities (\$)} \\
\toprule
 house & 750,000 & mortgage & 500,000 \\
 car & 30,000 & car loan & 20,000 \\
 bank deposit & 10,500 & credit card & 1,500 \\
 savings & 28,000 & &  \\
 bitcoin & 1,000 & net worth & 298,000 \\
\bottomrule
\end{tabular}
\quad
\begin{tabular}{ lr | lr } 
\multicolumn4c{\bf Alice's Balance Sheet (after)} \\
\toprule
\multicolumn2c{Assets (\$)} & \multicolumn2c{Liabilities (\$)} \\
\toprule
 house & 750,000 & mortgage & 500,000 \\
 car & 30,000 & car loan & 20,000 \\
 bank deposit & {\bf 1,500} & credit card & 1,500 \\
 savings & 28,000 & & \\
bitcoin & {\bf 11,000} & net worth & {\bf 299,000} \\
\bottomrule
\end{tabular}
\caption{Alice's balance sheet before (left) and after (right) mining \$10,000 worth of Bitcoin at an operational cost of \$9,000.}
\label{bitcoin1}
\end{table}

In Table \ref{bitcoin2} we see the effect of Alice using Bitcoin to purchase artwork from Bob. Assuming that no fees are paid, the net worth of each agent is not affected by the transaction, which corresponds to a pure exchange of assets. Notice again that, in this example, artwork is a type of asset similar to Bitcoin, in the sense that it is not a liability for any economic agent. Next in Table \ref{bitcoin3} we see the effect of Bob selling his bitcoins to his bank in exchange for an increase in his deposit account--- which is a liability for his bank. Assuming again that there are no fees for this transaction, it results in no change in net worth for either party. 

\begin{table}
\centering
\begin{tabular}{ lr | lr } 
\multicolumn4c{\bf Alice's Balance Sheet (before)} \\
\toprule
\multicolumn2c{Assets (\$)} & \multicolumn2c{Liabilities (\$)} \\
\toprule
 house & 750,000 & mortgage & 500,000 \\
 car & 30,000 & car loan & 20,000 \\
 bank deposit & 1,500 & credit card & 1,500 \\
 savings & 28,000 & & \\
bitcoin & 11,000 & & \\
artwork & 0  & net worth & 299,000 \\
\bottomrule
\end{tabular}
\quad
\begin{tabular}{ lr | lr } 
\multicolumn4c{\bf Alice's Balance Sheet (after)} \\
\toprule
\multicolumn2c{Assets (\$)} & \multicolumn2c{Liabilities (\$)} \\
\toprule
 house & 750,000 & mortgage & 500,000 \\
 car & 30,000 & car loan & 20,000 \\
 bank deposit & 1,500 & credit card & 1,500 \\
 savings & 28,000 & & \\
bitcoin & {\bf 3,000} &  & \\
artwork & {\bf 8,000} & net worth & 299,000 \\
\bottomrule
\end{tabular}
\vskip 0.2in 
\begin{tabular}{ lr | lr } 
\multicolumn4c{\bf Bob's Balance Sheet (before)} \\
\toprule
\multicolumn2c{Assets (\$)} & \multicolumn2c{Liabilities (\$)} \\
\toprule
 house & 1,200,000 & mortgage & 200,000 \\
 artwork & 80,000 & bank loan & 40,000 \\
 bank deposit & 10,000 & credit card & 500 \\
 savings & 150,000 & & \\
 bitcoin & 0 & net worth & 1,199,500 \\
\bottomrule
\end{tabular}
\quad
\begin{tabular}{ lr | lr } 
\multicolumn4c{\bf Bob's Balance Sheet (after)}  \\
\toprule
\multicolumn2c{Assets (\$)} & \multicolumn2c{Liabilities (\$)} \\
\toprule
 house & 1,200,000 & mortgage & 200,000 \\
 artwork & {\bf 72,000} & bank loan & 40,000 \\
 bank deposit & 10,000 & credit card & 500 \\
 savings & 150,000 & & \\
 bitcoin & {\bf 8,000} &  net worth & 1,199,500 \\ 
\bottomrule
\end{tabular}
\caption{Balance sheets before (left) and after (right) Alice uses \$8,000 in Bitcoin to purchase artwork from Bob.}
\label{bitcoin2}
\end{table}

\begin{table}
\centering
\begin{tabular}{ lr | lr } 
\multicolumn4c{\bf Bob's Balance Sheet (before)} \\
\toprule
\multicolumn2c{Assets (\$)} & \multicolumn2c{Liabilities (\$)} \\
\toprule
 house & 1,200,000 & mortgage & 200,000 \\
 artwork & 72,000 & bank loan & 40,000 \\
 bank deposit & 10,000 & credit card & 500 \\
 savings & 150,000 & & \\
 bitcoin & 8,000 & net worth & 1,199,500 \\
\bottomrule
\end{tabular}
\quad
\begin{tabular}{ lr | lr } 
\multicolumn4c{\bf Bob's Balance Sheet (after)} \\
\toprule
\multicolumn2c{Assets (\$)} & \multicolumn2c{Liabilities (\$)} \\
\toprule
 house & 1,200,000 & mortgage & 200,000 \\
 artwork & 72,000 & bank loan & 40,000 \\
 bank deposit & {\bf 17,000} & credit card & 500 \\
 savings & 150,000 & & \\
 bitcoin & {\bf 1,000} &  net worth & 1,199,500 \\
\bottomrule
\end{tabular}\quad
\vskip 0.2in 
\begin{tabular}{ lr | lr } 
\multicolumn4c{\bf Bank Balance Sheet (before)} \\
\toprule
\multicolumn2c{Assets (thousands \$)} & \multicolumn2c{Liabilities (thousands \$)} \\
\toprule
loans & 9,000,000 & deposits & 8,000,000 \\
treasuries  & 3,000,000 & savings  & 2,000,000 \\
reserves  & 300,000 & bonds  & 1,000,000 \\
cash  & 200,000 & & \\
bitcoin & 400,000 & net worth & 1,900,000 \\
\bottomrule
\end{tabular}
\quad
\begin{tabular}{ lr | lr } 
\multicolumn4c{\bf Bank Balance Sheet (after)} \\
\toprule
\multicolumn2c{Assets (thousands \$)} & \multicolumn2c{Liabilities (thousands \$)} \\
\toprule
loans & 9,000,000 & deposits & {\bf 8,000,007} \\
treasuries  & 3,000,000 & savings  & 2,000,000 \\
reserves  & 300,000 & bonds  & 1,000,000 \\
cash  & 200,000 & & \\
bitcoin & {\bf 400,007} & net worth & 1,900,000 \\
 \bottomrule
\end{tabular}
\caption{Balance sheets before (left) and after (right) Bob sells \$7,000 in Bitcoin to his bank.}
\label{bitcoin3}
\end{table}

The Bitcoin protocol originated an era of great expectations. Bitcoin
partisans think that the protocol can remove money from centralized
government control, eventually replace national currencies with BTCs,
decentralize payments, and provide anonymity to its users.

Bitcoin detractors point out that the protocol has a meager
transaction-per-second (TpS) capacity (3--6 TpS, while Visa, the credit card processor, processes 20 000
TpS) and high transaction fees, preventing everyday usage of BTC for
payments, is not sufficiently decentralized due to the natural tendency of
miners to coalesce, and does not provide anonymity. In addition, since BTC
has no intrinsic value, it can have any price, which is determined solely by
the supply and demand consideration, thus making its price extremely
volatile. Besides, as was mentioned earlier, the PoW consensus
algorithm results in enormous electricity consumption by the protocol for
securing a relatively low number of transactions (about 100 million per
year).

Bitcoin realists point out that while BTC cannot be used for conventional
payments, it can be used for the immutable and final discharge of
substantial obligations. The protocol is not nearly as decentralized as was
initially claimed but still inspires future developments. The lack of
anonymity can be addressed if necessary.

\begin{figure}[h!]
\centering
   \includegraphics[trim=0 7cm 0 7cm,width=\textwidth]{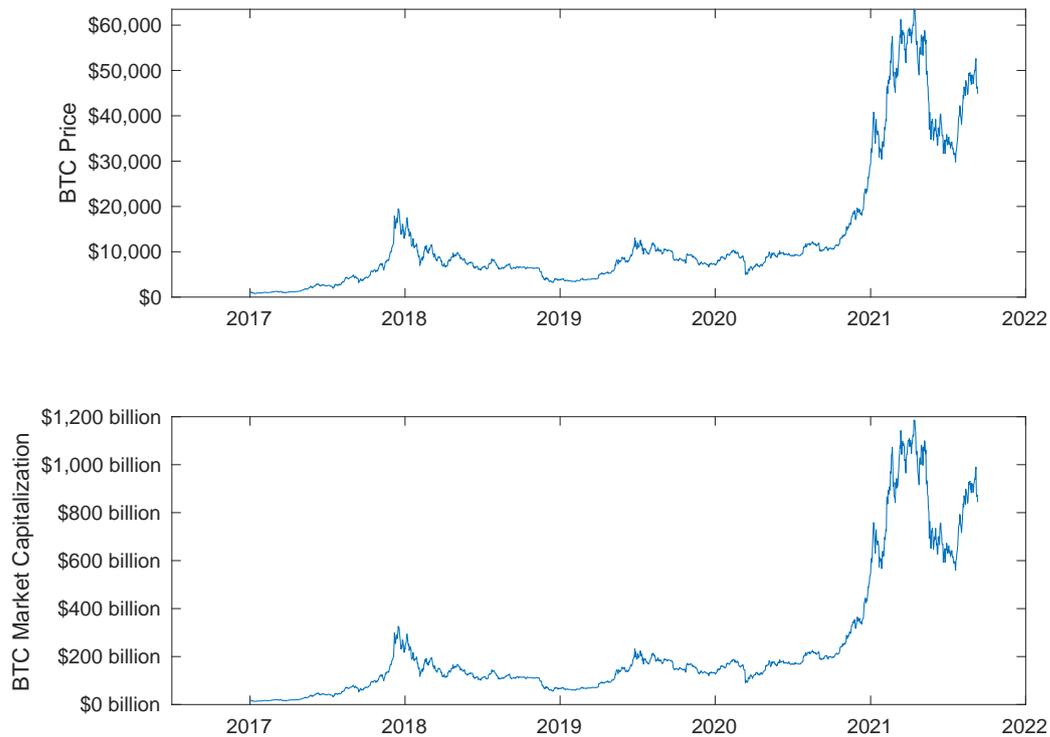}
 \caption{Price (top) and market capitalization (bottom) for Bitcoin (BTC) in USD. Source: coinmarketcap.com, downloaded on Sept 11, 2021.}
 \label{bitcoin_figure}
\end{figure}

Figure \ref{bitcoin_figure} shows the price and market capitalization of BTC over the last five years, a period marked by both extraordinary growth and extreme volatile swings. Observe that, because of the slow process of mining new coins, the market capitalization behaves for practical purposes as a near-constant multiple of the price, and this relationship tends to become more exact as the number of BTC in circulation approach its asymptotic limit of 21 million coins\footnote{As of September 2021, there were close to 18.8 million BTC in circulation, of which 10 to 20\% might have been irrecoverably lost.}. 

Speaking of Helen of Troy, Christopher Marlowe coined an immortal phrase: ``the face that launched a thousand ships." Without a doubt, Bitcoin can be called ``the protocol that launched a thousand coins." Many associated protocols are modest variations of the original one. Ethereum, however, is genuinely different. 

Launched in 2015, Ethereum promises were even grander than Bitcoin's, and with a current market capitalization of approximately 450 Billion USD, it is Bitcoin's closest rival in popularity. From the very beginning \cite{Buterin2013}, it was designed as a decentralized virtual machine, called Ethereum Virtual Machine (EVM), i.e., the first distributed Turing-complete computer. EVM can process smart contracts, support distributed autonomous organizations, thus providing consensus as a service (CaaS).   

In reality, Ethereum suffers from several drawbacks (see \cite[Chapter 6]{LiptonTreccani2021} for more details). First, it has low TpS capacity and high transaction fees, preventing its usage for everyday transactions. Second, an obsolete payment model based on gas consumption makes EVM a distributed calculator at best. Third, smart contracts are not smart enough, cannot be fixed if they have bugs, and consume collateral on a prodigious scale. Finally, Ethereum also uses a PoW consensus algorithm and is a voracious consumer of electricity. And yet, as a robust CaaS provider, EVM is very convenient for issuing tokens, such as the stable coins discussed below, and for many other purposes. Moreover, it is a veritable triumph of engineering. Ethereum launched the CaaS revolution; presently, it competes with several newer protocols with superior capabilities, such as Cardano, Polkadot, Solana, Algorand, and Zilliqa.

\begin{figure}[h!]
\centering
   \includegraphics[trim=0 7cm 0 7cm,width=\textwidth]{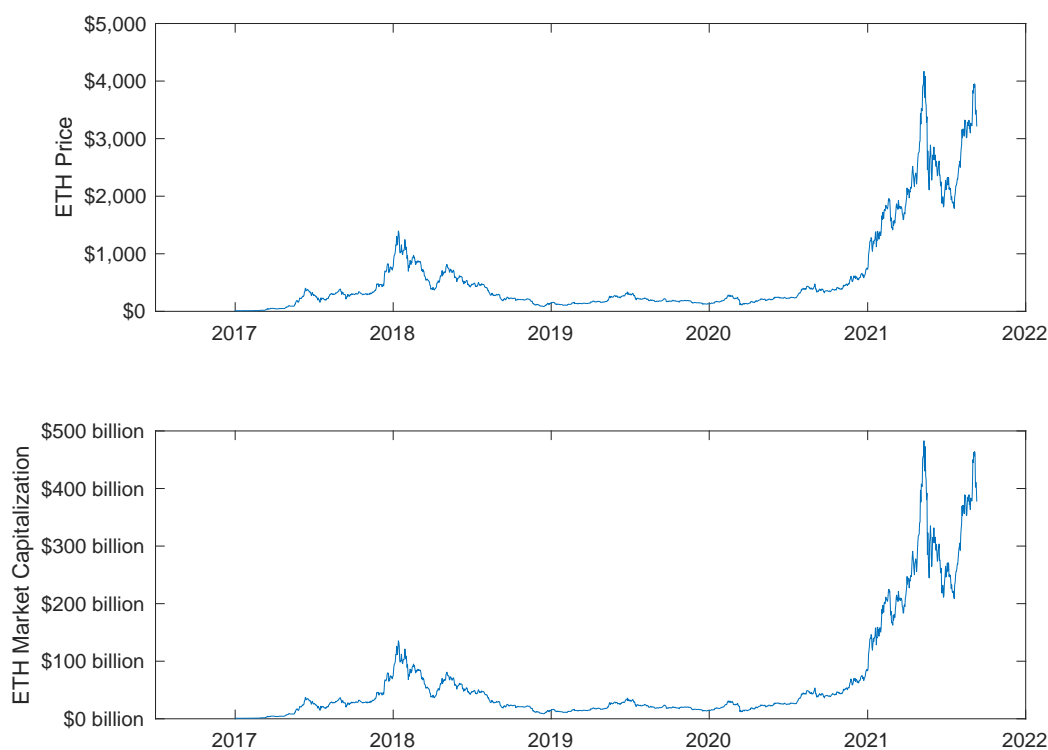}
 \caption{Price (top) and market capitalization (bottom) for Ethereum (ETH) in USD. Source: coinmarketcap.com, downloaded on Sept 11, 2021.}
 \label{ETH_figure}
\end{figure}

Figure \ref{ETH_figure} shows the price and market capitalization of Ethereum's native token ETH, which has exhibited a similar pattern of rapid growth and extreme volatility as BTC. Unlike BTC, the number of ETH coins in circulation was not designed to converge to a fixed asymptotic value, and has increased more or less linearly from the initial 72 million coins to currently close to 120 million coins, although this rate of increase is also subject to change by consensus within the network. 

The third prominent example we cover in this class is Ripple. Even though, as we explained above, Bitcoin, Ethereum, and Ripple all represent pure assets of the holder, they do have substantial, if subtle, differences. At a technical level, the main difference is that, in the Ripple protocol, consensus in the ledger is achieved by voting among a special type of network nodes called validators (see \cite[Chapter 7]{LiptonTreccani2021} for details). That is to say, the time and energy-consuming consensus by PoW used by the Bitcoin and Ethereum protocols are replaced by a much faster process, at the expense of a higher degree of centralization. Relatedly, from an economic point of view, the key difference is that all XRP coins---the native tokens in the protocol---were pre-mined at the inception of the network, rather than mined overtime as reward for validators as is the case with Bitcoin and Ethereum. Concretely, 100 billion XRP coins were created in 2012, with 80 billion coins allocated to the company behind the protocol and 20 billion coins to its founders, and have been put in circulation at a more or gradual pace of approximately 1 billion coins per month through spending by the original owners (that is to say, exchanging them by other assets). 

\begin{figure}[h!]
\centering
   \includegraphics[trim=0 7cm 0 7cm,width=\textwidth]{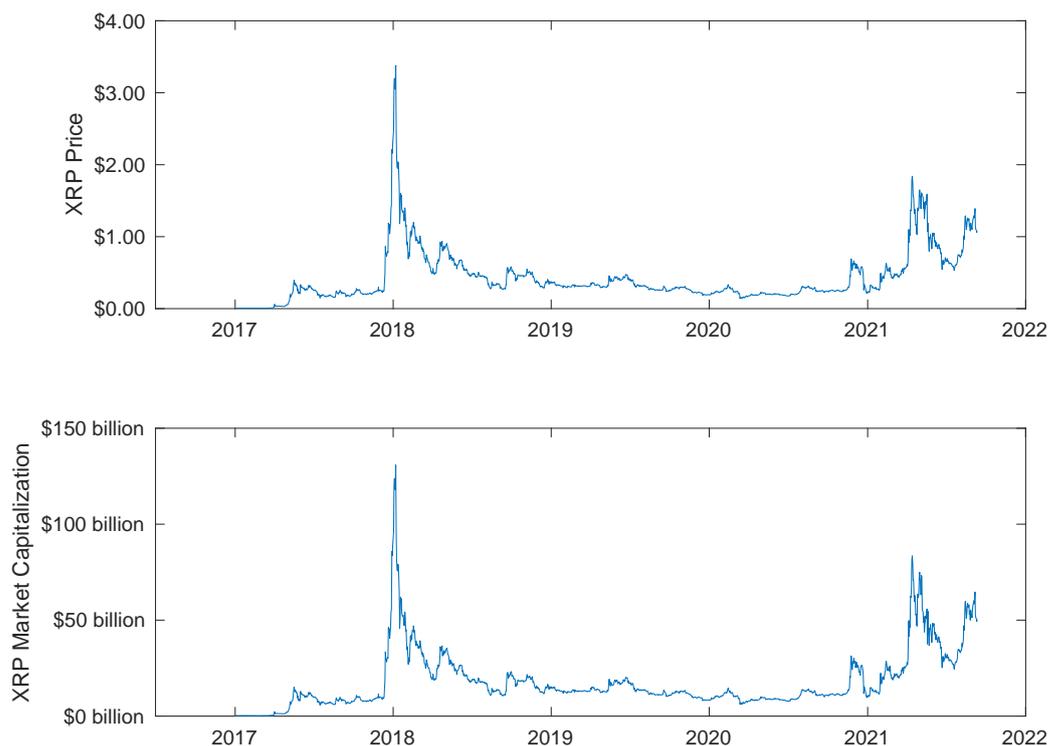}
 \caption{Price (top) and market capitalization (bottom) for Ripple (XRP) in USD. Source: coinmarketcap.com, downloaded on Sept 12, 2021.}
 \label{ripple_figure}
\end{figure}

Because of its faster and cheaper validation of transactions, Ripple became a popular protocol for large cross-border payments, primarily as a competitor to traditional payment systems such as SWIFT. As can be seen in Figure \ref{ripple_figure}, the price of XRP has exhibited at least as much volatility as that of Bitcoin and Ethereum, essentially for the same reason, that is to say, that there is no other asset, financial or real, backing its value. In addition, while Bitcoin, was deemed by Security and Exchange Commission (SEC) Chair Jay Clayton as not being a security\footnote{``Cryptocurrencies are replacements for sovereign currencies… ... That type of currency is not a security.'' See \url{https://www.cnbc.com/amp/2018/06/06/sec-chairman-clayton-says-agency-wont-change-definition-of-a-security.html?}.}, the SEC filed a lawsuit against Ripple, alleging that by selling XRP, the company held a \$1.3 billion unregistered securities offering\footnote{See \url{https://www.sec.gov/news/press-release/2020-338}.}.

\section{Central bank digital currencies (CBDC)}
\label{CBDC_section}

In order of economic complexity, the next class of cryptocurrency consists of digital coins issued by a central bank, the so-called central bank digital currency (CBDC). Several countries, including the Bahamas, China, Ecuador, the European Union, and Switzerland, to mention but a few, are currently experiment with CBDCs \cite{BoarWehrli2021}. The US and UK are less active in this area. Economically speaking, cryptocurrencies of this type are most similar to cash and central bank reserves. 

Recall that central bank reserves are special types of deposit accounts that commercial banks (and other select financial institutions) maintain with the central bank. They are therefore assets for the commercial bank and liabilities for the central bank, and are the primary instruments used to conduct monetary policy. Although completely digital, reserves should not, in our view, be considered a cryptocurrency because of their limited access and centralized, non-anonymous nature: at any give time, the central bank knows exactly the amount of reserves held by each participant bank, as well as any transfers occurring between banks. 

The other main type of central bank liability is physical cash, either in the form of notes or coins. Unlike reserves, cash can be held as an asset by any economic agent. Moreover, with the exception of tracing mechanisms used in criminal investigations, it is completely anonymous, in the sense that the central bank only knows the total amount of cash in circulation, but does not know how much is held by which agent or what transactions are made between agents using cash. 

Analogously to cash and reserves, CBDC are assets for the economic agents holding them and liabilities for the central bank and can therefore, in principle, be used for the same economic purposes. If  restricted to banks only, they would be indistinguishable from reserves, so the only interesting situation consist of CBDC that can be held by the general public. One possibility consist of CBDC in the form of generalized reserves, that is to say, deposit accounts that members of the public hold at the central bank (see \cite{BordoLevin2017}). This would not require any technological innovation, other than scaling up the system of accounts already used by commercial banks and select financial institutions. It would {\em not}, however, have the property of anonymity: as it is the case with reserves, the central bank would know the amount of CBDC held by each economic agent and be able to trace every transaction within the system. Moreover, by offering accounts to all members of the public, a central bank would have to perform the Know Your Client (KYC) and Anti Money Laundering (AML) functions currently performed by commercial banks, for which it is most likely ill equipped. 

A much more promising possibility consists of a central bank partnering with private institutions to create what the Bank of England has termed a {\em platform } model for CBDC (see \cite{BOE2020}). In this model, the central bank maintains a core ledger recording only basic payment transactions between accounts that are accessible only to private sector firms known as Payment Interface Providers, which would be responsible for all interactions with the general public, from web interfaces to value-added services. The Payment Interface Provider would either register an account in the core ledger for each of its clients, or maintain a single pooled account for all of its clients. Crucially, in this platform model, the identity of the users does not need to be known to the central bank, as the accounts could be pseudonymous on the core ledger. For the sake of argument, suppose that the Payment Interface Provider is a commercial bank. When a member of the public who is a client of the bank requests to convert part of their bank deposits into CBDC, digital tokens are created in the core ledger in the form of a liability for the central bank and are held as assets for the client in a digital wallet administered by the bank, while at the same time an equivalent amount is deducted from the bank's reserves. From a balance sheet perspective, this is entirely analogous to a cash withdraw, whereby there is a decrease in the amount of deposits and a corresponding decrease in the amount of cash/reserves held by the bank. The only difference is that, unlike cash and real wallets, CBDC and the digital wallets in which they are stored are still administered by the Payment Interface Provider, in this case the bank.    

When properly designed, this type of CBDC can have all the properties commonly associated with cash, including anonymity with respect to the central bank, except in extraordinary circumstances such as criminal investigations. Importantly, the KYC and AML functions would continue to be performed by Payment Interface Provider. Finally, the central bank would guarantee that CBDC tokens held by commercial banks can always be converted into cash at par, thereby ensuring the stability of the CBDC price with respect to the fiat currency issued by the same central bank. A novel and more controversial use of CBDC could be the implementation of unconventional monetary policy in the form of negative interest rates, whereby wholesale replacement of physical by digital cash would remove, or at least relax, the so called zero lower bound (ZLB) (see \cite{GrasselliLipton2019a} for details). 

Table \ref{CBDC1} shows illustrates the balance sheet operations described above. Namely, Bob uses the same commercial bank as in the previous example to acquire CDBC. From Bob's point of view, this is an exchange of assets (deposits replaced by CBDC), whereas from the bank's point of view it is a decrease in both assets (reserves) and liabilities (deposits), with neither party experiencing any change in net worth. Finally, from the point of view of the central bank, this operation is an exchange of liabilities (reserves replaced by CBDC). Observe also that, as shown in the example, CBDC can co-exist with Bitcoins, bank deposits and other types of assets.

\begin{table}
\centering
\begin{tabular}{ lr | lr } 
\multicolumn4c{\bf Bob's Balance Sheet (before)} \\
\toprule
\multicolumn2c{Assets (\$)} & \multicolumn2c{Liabilities (\$)} \\
\toprule
 house & 1,200,000 & mortgage & 200,000 \\
 artwork & 72,000 & bank loan & 40,000 \\
 bank deposit & 17,000 & credit card & 500 \\
 savings & 150,000 & & \\
 bitcoin & 1,000 & & \\
 CBDC & 0 & net worth & 1,191,500 \\
\bottomrule
\end{tabular}
\quad
\begin{tabular}{ lr | lr } 
\multicolumn4c{\bf Bob's Balance Sheet (after)} \\
\toprule
\multicolumn2c{Assets (\$)} & \multicolumn2c{Liabilities (\$)} \\
\toprule
 house & 1,200,000 & mortgage & 200,000 \\
 artwork & 72,000 & bank loan & 40,000 \\
 bank deposit & {\bf 11,000} & credit card & 500 \\
 savings & 150,000 & & \\
 bitcoin & 1,000 &  & \\
 CBDC & {\bf 6,000} & net worth & 1,191,500 \\
\bottomrule
\end{tabular}\quad
\vskip 0.2in 
\begin{tabular}{ lr | lr } 
\multicolumn4c{\bf Bank Balance Sheet (before)} \\
\toprule
\multicolumn2c{Assets (thousands \$)} & \multicolumn2c{Liabilities (thousands \$)} \\
\toprule
loans & 9,000,000 & deposits & 8,000,007 \\
treasuries  & 3,000,000 & savings  & 2,000,000 \\
reserves  & 300,000 & bonds  & 1,000,000 \\
cash  & 200,000 & & \\
bitcoin & 400,007 & & \\
& & net worth & 1,900,000 \\
\bottomrule
\end{tabular}
\quad
\begin{tabular}{ lr | lr } 
\multicolumn4c{\bf Bank Balance Sheet (after)} \\
\toprule
\multicolumn2c{Assets (thousands \$)} & \multicolumn2c{Liabilities (thousands \$)} \\
\toprule
loans & 9,000,000 & deposits & {\bf 8,000,001} \\
treasuries  & 3,000,000 & savings  & 2,000,000 \\
reserves  & {\bf 299,994} & bonds  & 1,000,000 \\
cash  & 200,000 & & \\
bitcoin & 400,007 & & \\
& & net worth & 1,900,000 \\
\bottomrule
\end{tabular}

\vskip 0.2in  
\begin{tabular}{ lr | lr } 
\multicolumn4c{\bf Central Bank Balance Sheet (before)} \\
\toprule
\multicolumn2c{Assets (thousands \$)} & \multicolumn2c{Liabilities (thousands \$)} \\
\toprule
treasuries  & 390,000,000 & reserves  & 180,000,000 \\
& & cash  & 100,000,000 \\
& & CBDC  & 60,000,000 \\
 & & net worth  & 50,000,000 \\
\bottomrule
\end{tabular}
\quad
\begin{tabular}{ lr | lr } 
\multicolumn4c{\bf  Central Bank Balance Sheet (after)} \\
\toprule
\multicolumn2c{Assets} & \multicolumn2c{Liabilities} \\
\toprule
 treasuries  & 390,000,000 & reserves  & {\bf 179,999,994} \\
& & cash  & 100,000,000 \\
& & CBDC  & {\bf 60,000,006} \\
 & & net worth  & 50,000,000 \\
\bottomrule
\end{tabular}\quad
\caption{Balance sheets before (left) and after (right) Bob buys \$6,000 in central bank digital currency (CBDC) from his Bank.}
\label{CBDC1}
\end{table}

\section{Stable Coins}
\label{stable_section}

Stable coins are a class of cryptocurrency designed to avoid the wild price fluctuations that plagued pure-asset coins such as Bitcoins and Ethereum since their inception (see Figures \ref{bitcoin_figure} and \ref{ETH_figure} above). From an economic point of view, their key feature is that they are a liability of some financial institution that is obligated to hold an equal of larger amount of some other asset as collateral (see \cite{LiptonSardonScharSchupbach2021} for more details). 

\subsection{Fiat-backed stable coins (FBSC)} 

The simplest example of stable coin consists of a cryptocurrency issued by a bank or some other private financial institutions that is committed to hold an equal or greater amount of central bank reserves\footnote{Or, equivalently, cash or CBDC, given the central bank commitment to par convertibility between cash, CBDC, and reserves.} as assets. Similar to the case of a CBDC, when a client of such bank decides to convert part of their bank deposits into the FBSC, they receive digital tokens in the form of assets held in their digital wallet, and consequently the proposed implementations for these two types of cryptocurrencies are very similar from a technological point of view. As a matter of fact, some authors considered FBSC as {\em synthetic} CBDC (see \cite{AdrianMacriniGriffoli2019}). 

From an economic point of view, the key difference between the two is that FBSC are a financial liability of the private company issuing them, whereas CBDC, as discussed above, are a liability of the central bank. In particular, FBSC can be introduced by private companies even if the corresponding central bank is unwilling or unable to issue CBDC, provided the issuer has enough cash or central bank reserves to fully collateralize the coins. 

In the case of a bank issuing FBSC, an obvious difficulty arises when the amount of coins demanded by the clients, which in principle can be as high as their total deposits, exceeds the amount of reserves held by the bank. In this case, in order to credibly maintain the promise of convertibility, the bank must sell some of its other assets in order to raise the necessary amount of reserves as collateral, which might not always be possible. This is entirely analogous to the liquidity problem that arises when bank clients demand to withdraw deposits in excess of reserves, and the only guaranteed solution to this problem is to make sure that an emitter of FBSC always has reserves in excess of deposits. In other words, the natural emitter of a FBSC is what is known as a narrow bank. 

An example of FBSC currently in circulation is the Utility Settlement Coin (USC) designed by the fintech company Clearmatics. 
These coins are meant to be used as an internal token for interbank payments between participating banks, and are fully collateralized by their collective 
central bank reserves (see \cite[Chapter 8]{LiptonTreccani2021} for details). As mentioned above, in order for circulation of these coins to be extended to the general public, their issuance will have to be outsourced to a narrow bank. 

Table \ref{FBSC1} illustrates the operation of client of a narrow bank, in this case Alice from the previous examples, converting part of her bank deposits into FBSC. Notice that the narrow bank in this example is similar in size (for example as measured by total assets) to the conventional bank depicted in Table \ref{CBDC1}, but with a much narrower set of assets (hence the name), namely primarily central bank liabilities (i.e reserves, cash, and CBDC). This does not prevent a narrow bank from lending (i.e holding regular loans as assets), provided they have other types of liabilities to fund these lending activities, such as savings (e.g 2-year savings accounts) in this example. The crucial feature is that they hold reserves in excess of the amount of FBSC and deposits. Notice further that, because there is no risk of a run on deposits, a narrow bank can operate with a much slimmer amount of capital\footnote{See \cite{GrasselliLipton2019b} for an example of a macroeconomic model for narrow banking.}.

\begin{table}
\centering
\begin{tabular}{ lr | lr } 
\multicolumn4c{\bf Alice's Balance Sheet (before)} \\
\toprule
\multicolumn2c{Assets (\$)} & \multicolumn2c{Liabilities (\$)} \\
\toprule
 house & 750,000 & mortgage & 500,000 \\
 car & 30,000 & car loan & 20,000 \\
 bank deposit & 1,500 & credit card & 1,500 \\
 savings & 28,000 & & \\
bitcoin & 3,000 &  & \\
artwork & 8,000 & & \\
FBSC & 0 & net worth & 299,000 \\
\bottomrule
\end{tabular}
\quad
\begin{tabular}{ lr | lr } 
\multicolumn4c{\bf Alice's Balance Sheet (after)} \\
\toprule
\multicolumn2c{Assets (\$)} & \multicolumn2c{Liabilities (\$)} \\
\toprule
 house & 750,000 & mortgage & 500,000 \\
 car & 30,000 & car loan & 20,000 \\
 bank deposit & {\bf 500} & credit card & 1,500 \\
 savings & 28,000 & & \\
bitcoin & 3,000 &  & \\
artwork & 8,000 & & \\
FBSC & {\bf 1,000} & net worth & 299,000 \\
\bottomrule
\end{tabular}
\vskip 0.2in 
\begin{tabular}{ lr | lr } 
\multicolumn4c{\bf Narrow Bank Balance Sheet (before)} \\
\toprule
\multicolumn2c{Assets (thousands \$)} & \multicolumn2c{Liabilities (thousands \$)} \\
\toprule
loans & 2,500,000 & savings  & 2,500,000 \\
reserves  & 11,000,000 & deposits & 10,000,00 \\
cash  & 100,000 & FBSC  & 1,000,000 \\
CBDC & 150,000 & & \\
 & & net worth & 250,000 \\
\bottomrule
\end{tabular}
\quad
\begin{tabular}{ lr | lr } 
\multicolumn4c{\bf Narrow Bank Balance Sheet (after)} \\
\toprule
\multicolumn2c{Assets (thousands \$)} & \multicolumn2c{Liabilities (thousands \$)} \\
\toprule
loans & 2,500,000 & savings  & 2,500,000 \\
reserves  & 11,000,000 & deposits & {\bf 9,999,999} \\
cash  & 100,000 & FBSC  & {\bf 1,000,001} \\
CBDC & 150,000 & & \\
 & & net worth & 250,000 \\
 \bottomrule
\end{tabular}
\caption{Balance sheets before (left) and after (right) Alice buys \$1,000 in fiat-backed stable coins (FBSC) from her Narrow Bank.}
\label{FBSC1}
\end{table}

\subsection{Custodial stable coins (CSC)}

These are similar to FBSC, except that, instead of central bank reserves, the issuer is obligated to hold deposits in a designated bank as a collateral. In this case, price stability is achieved, in principle, because the issuer guarantees convertibility of the CSC into cash by being able to withdraw from the deposit account held as collateral. Accordingly, this relies on the capacity of the designated bank to honour the deposit withdraw, which in turn depends on the designated bank having access to enough central bank reserves. That is to say, similarly to FBSC, price stability of CSC can only be fully guaranteed if the designated bank is itself a narrow bank. 

\begin{figure}[h!]
\centering
   \includegraphics[trim=0 7cm 0 7cm,width=\textwidth]{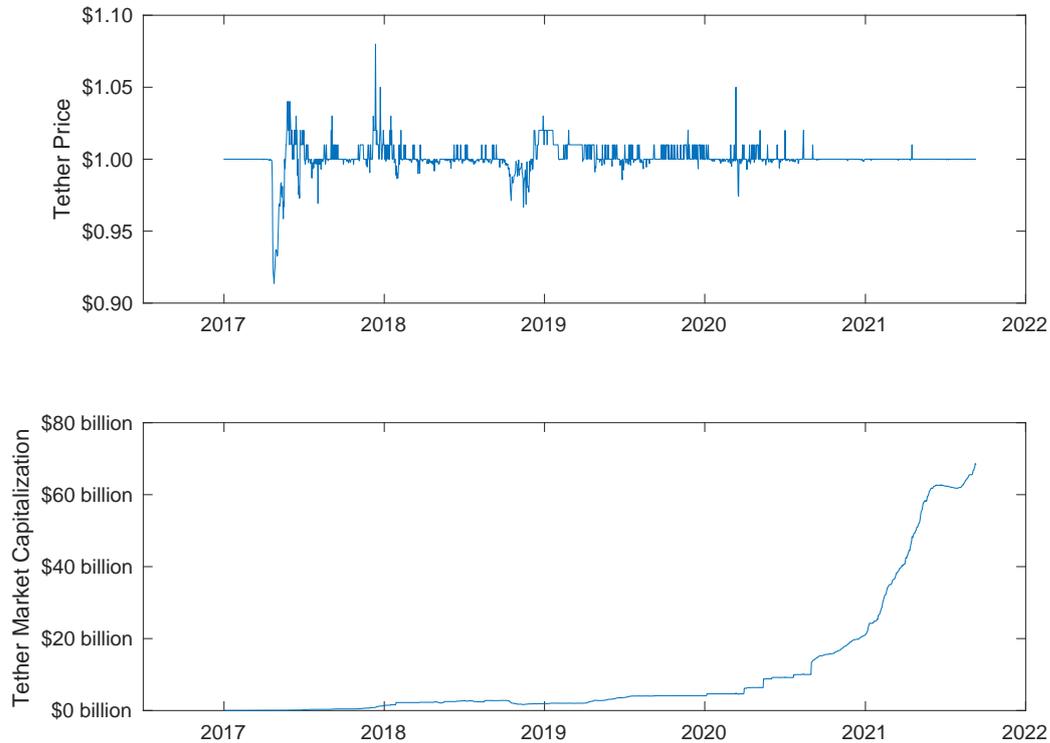}
 \caption{Price (top) and market capitalization (bottom) for Tether (USDT) in USD. Source: coinmarketcap.com, downloaded on Sept 11, 2021.}
 \label{tether_figure}
\end{figure}

Current examples of CSC include Tether (USDT), TrueUSD (TUSD), USD Coin (USDC) and SilaToken (SILA). Tether is by far the best known example in this class, and the fact that its price {\em does} fluctuate agains the USD (see Figure \ref{tether_figure}) indicates that, in practice, there is still a lot of opacity around the deposits used as collateral. 
Fluctuations highlight that privately issued CSC always carries some credit risk, mainly because the bank where it holds collateral can default or collateral itself can be absent. Under normal conditions, banks, being tightly regulated institutions, default very infrequently. However, during times of crisis, their creditworthiness can deteriorate rapidly, thus jeopardizing the very stability of the corresponding CSCs.

Table \ref{CSC1} illustrates the operation of a member of the public, in this case Bob from the previous examples, converting part of his bank deposits into a CSC 
issued by a generic Custodial company. We assume the custodial company keeps deposits in the same narrow bank used by Alice in the previous example. The changes in balance sheet that accompany this operation are as follows: Bob has his bank deposits reduced by the same amount of CSC purchased; consequently, the deposit liabilities of Bob's bank are reduced by this amount, with a corresponding transfer of reserves to the narrow bank; the deposit account of the custodial company increases by this amount, which allows it to issue the new CSC. Notice that the narrow bank has enough reserves to cover not only the FBSC that it issues, but also all of its deposits, which includes the deposit account of the Custodial company, thereby ensuring the price stability of the CSC. If the Custodial company used the same bank as Bob instead, it would might run into liquidity issues if it attempted to convert all of its bank deposits into cash, which would put into question the convertibility of its own CSC. 

\begin{table}
\centering
\begin{tabular}{ lr | lr } 
\multicolumn4c{\bf Bob's Balance Sheet (before)} \\
\toprule
\multicolumn2c{Assets (\$)} & \multicolumn2c{Liabilities (\$)} \\
\toprule
 house & 1,200,000 & mortgage & 200,000 \\
 artwork & 72,000 & bank loan & 40,000 \\
 bank deposit & 11,000 & credit card & 500 \\
 savings & 150,000 & & \\
 bitcoin & 1,000 &  & \\
 CBDC & 6,000 & & \\
 CSC & 0 & net worth & 1,199,500 \\
\bottomrule
\end{tabular}
\quad
\begin{tabular}{ lr | lr } 
\multicolumn4c{\bf Bob's Balance Sheet (after)} \\
\toprule
\multicolumn2c{Assets (\$)} & \multicolumn2c{Liabilities (\$)} \\
\toprule
 house & 1,200,000 & mortgage & 200,000 \\
 artwork & 72,000 & bank loan & 40,000 \\
 bank deposit & {\bf 9,000} & credit card & 500 \\
 savings & 150,000 & & \\
 bitcoin & 1,000 &  & \\
 CBDC & 6,000 & & \\
 CSC & {\bf 2,000} & net worth & 1,199,500 \\
\bottomrule
\end{tabular}\quad
\vskip 0.2in 
\begin{tabular}{ lr | lr } 
\multicolumn4c{\bf Custodial Balance Sheet (before)} \\
\toprule
\multicolumn2c{Assets (thousands \$)} & \multicolumn2c{Liabilities (thousands \$)} \\
\toprule
bank deposits & 310,000 & CSC & 300,000 \\
 & & net worth & 10,000 \\
\bottomrule
\end{tabular}
\quad
\begin{tabular}{ lr | lr } 
\multicolumn4c{\bf Custodial Balance Sheet (after)} \\
\toprule
\multicolumn2c{Assets (thousands \$)} & \multicolumn2c{Liabilities (thousands \$)} \\
\toprule
bank deposits & 310,002 & CSC & 300,002 \\
 & & net worth & 10,000 \\ \bottomrule
\end{tabular}
\vskip 0.2in 
\begin{tabular}{ lr | lr } 
\multicolumn4c{\bf Bank Balance Sheet (before)} \\
\toprule
\multicolumn2c{Assets (thousands \$)} & \multicolumn2c{Liabilities (thousands \$)} \\
\toprule
loans & 9,000,000 & deposits & 8,000,001 \\
treasuries  & 3,000,000 & savings  & 2,000,000 \\
reserves  & 150,000 & bonds  & 1,000,000 \\
cash  & 200,000 & & \\
bitcoin & 400,007 &  \\
CBDC & 149,994 & net worth & 1,900,000 \\
\bottomrule
\end{tabular}
\quad
\begin{tabular}{ lr | lr } 
\multicolumn4c{\bf Bank Balance Sheet (after)} \\
\toprule
\multicolumn2c{Assets (thousands \$)} & \multicolumn2c{Liabilities (thousands \$)} \\
\toprule
loans & 9,000,000 & deposits & {\bf 7,999,999} \\
treasuries  & 3,000,000 & savings  & 2,000,000 \\
reserves  & {\bf 149,998} & bonds  & 1,000,000 \\
cash  & 200,000 & & \\
bitcoin & 400,007 &  \\
CBDC & 149,994 & net worth & 1,900,000 \\
 \bottomrule
\end{tabular}
\vskip 0.2in 
\begin{tabular}{ lr | lr } 
\multicolumn4c{\bf Narrow Bank Balance Sheet (before)} \\
\toprule
\multicolumn2c{Assets (thousands \$)} & \multicolumn2c{Liabilities (thousands \$)} \\
\toprule
loans & 2,500,000 & savings  & 2,500,000 \\
reserves  & 11,000,000 & deposits & 9,999,999 \\
cash  & 100,000 & FBSC  & 1,000,001 \\
CBDC & 150,000 & & \\
 & & net worth & 250,000 \\
\bottomrule
\end{tabular}
\quad
\begin{tabular}{ lr | lr } 
\multicolumn4c{\bf Narrow Bank Balance Sheet (after)} \\
\toprule
\multicolumn2c{Assets (thousands \$)} & \multicolumn2c{Liabilities (thousands \$)} \\
\toprule
loans & 2,500,000 & savings  & 2,500,000 \\
reserves  & {\bf 11,000,002} & deposits & {\bf 10,000,001} \\
cash  & 100,000 & FBSC  & 1,000,001 \\
CBDC & 150,000 & & \\
 & & net worth & 250,000 \\
 \bottomrule
\end{tabular}

\caption{Balance sheets before (left) and after (right) Bob buys \$2,000 in custodial stable coin (CSC) from a Custodial Company.}
\label{CSC1}
\end{table}

The growth of Tether and other stable coins has been phenomenal. In one year, the capitalization of Tether increased by a factor of six: it was about \$10 billion on August 1st, 2020, and \$64 billion on August 1st, 2021 (see Figure \ref{tether_figure}). We can explain this growth by noticing that trading cryptocurrencies cannot rely on conventional banks because they cannot move funds in real-time and are disinclined to get deeply involved in crypto trading. Accordingly, it must use stable coins instead.  

\subsection{Digital trade coins (DTC)}

Next in the hierarchy of stable coins are cryptocurrencies collateralized with a basket of either real or financial assets. This class includes the generic digital trade coins (DTC) proposed in \cite{LiptonHardjonoPentland2018}, as well as several specific examples including Tiberiuscoin (TCX), a cryptocurrency collateralized by a basket of previous metals, and Diem (formerly known a Libra), the proposed Facebook cryptocurrency collateralized by a basket of financial assets including currencies and US Treasuries. 

The defining feature of DTC is that an administrator assembles a pool of underlying assets originally owned by a group of sponsors and issues DTC in an amount initially equal to the value of the collateral. The administrator then sells the DTC to the general public in exchange of fiat currency, which is held in a deposit account in an affiliated bank on behalf of the sponsors. The end result is: the administrator's balance sheet consists of the underlying basket as assets and the DTC as liabilities; the sponsors have deposits in the affiliated bank instead of the underlying basket of assets; and the general public, who originally held fiat currency, holds DTC as assets. 

This setup ensures that the price of the DTC remains close to the market price of the underlying basket. This is because if the price of the DTC falls significantly, then the general public can sell it back to the administrator, who in turn returns a portion of the underlying assets to the sponsors in exchange for fiat currency that can be used to pay back the general public. Observe that this is contingent on the affiliated bank having enough reserves to cover the withdraws made by the sponsors, so once more the natural candidate for such role is a narrow bank. Conversely, if the price of the DTC increases, it is in the interest of the sponsor to supply additional assets to the administrator, who then issues and sells additional DTC and passes the proceeds back to the sponsor in the form of deposits in the affiliated bank. Secondary mechanisms for price stability of the DTC include savings accounts held by the general public with variable interest rates paid in DTC according to changes in demand.   

Table \ref{DTC1} illustrates this last situation. A group of sponsors with a combined net worth of \$100,000 and \$400,000 in loans uses these funds to purchase a basket of assets worth \$500,000. They then keep \$50,000 of these assets in their own balance sheet, while transferring \$450,000 to the Administrator, who issues an equal amount of DTC and sells to the general public. The proceeds of the sales are held as deposits in a narrow bank, which are listed in the balance sheet of the sponsors as assets. When the price of the DTC increases, the administrator finds a member of the general public interested in buying additional DTC, in this case Bob from the previous examples, and the following operations happen: Bob's bank deposits go down, and consequently the deposit liabilities of his bank, with an equal amount of reserves being transferred to the narrow bank; the sponsors transfer the same amount from the basket of assets to the administrators in exchange for an increase in deposits with the narrow bank; the administrator issues new DTCs. 

\begin{table}
\centering
\begin{tabular}{ lr | lr } 
\multicolumn4c{\bf Bob's Balance Sheet (before)} \\
\toprule
\multicolumn2c{Assets (\$)} & \multicolumn2c{Liabilities (\$)} \\
\toprule
house & 1,200,000 & mortgage & 200,000 \\
 artwork & 72,000 & bank loan & 40,000 \\
 bank deposit & 9,000 & credit card & 500 \\
 savings & 150,000 & & \\
 bitcoin & 1,000 &  & \\
 CBDC & 6,000 & & \\
 CSC & 2,000 & & \\
 DTC & 0 & net worth & 1,199,500 \\
\bottomrule
\end{tabular}
\quad
\begin{tabular}{ lr | lr } 
\multicolumn4c{\bf Bob's Balance Sheet (after)} \\
\toprule
\multicolumn2c{Assets (\$)} & \multicolumn2c{Liabilities (\$)} \\
\toprule
 house & 1,200,000 & mortgage & 200,000 \\
 artwork & 72,000 & bank loan & 40,000 \\
 bank deposit & {\bf 4,000} & credit card & 500 \\
 savings & 150,000 & & \\
 bitcoin & 1,000 &  & \\
 CBDC & 6,000 & & \\
 CSC & 2,000 & & \\
 DTC & {\bf 5,000} & net worth & 1,191,500 \\
\bottomrule
\end{tabular}
\vskip 0.2in 
\begin{tabular}{ lr | lr } 
\multicolumn4c{\bf Administrator Balance Sheet (before)} \\
\toprule
\multicolumn2c{Assets (thousands \$)} & \multicolumn2c{Liabilities (thousands \$)} \\
\toprule
basket of assets & 450,000 & DTC & 450,000 \\
 & & net worth & 0 \\
\bottomrule
\end{tabular}
\quad
\begin{tabular}{ lr | lr } 
\multicolumn4c{\bf Administrator Balance Sheet (after)} \\
\toprule
\multicolumn2c{Assets (thousands \$)} & \multicolumn2c{Liabilities (thousands \$)} \\
\toprule
basket of assets & {\bf 450,005} & DTC & {\bf 450,005} \\
 & & net worth & 0 \\
 \bottomrule
\end{tabular}
\vskip 0.2in 
\begin{tabular}{ lr | lr } 
\multicolumn4c{\bf Sponsors Balance Sheet (before)} \\
\toprule
\multicolumn2c{Assets (thousands \$)} & \multicolumn2c{Liabilities (thousands \$)} \\
\toprule
basket of assets & 50,000 & loans & 400,000 \\
deposits  & 450,000 & & \\
& & net worth & 100,000 \\
\bottomrule
\end{tabular}
\quad
\begin{tabular}{ lr | lr } 
\multicolumn4c{\bf Sponsors Balance Sheet (after)} \\
\toprule
\multicolumn2c{Assets (thousands \$)} & \multicolumn2c{Liabilities (thousands \$)} \\
\toprule
basket of assets & {\bf 49,995} & loans & 400,000 \\
deposits  & {\bf 450,005} & & \\
& & net worth & 100,000 \\
 \bottomrule
\end{tabular}
\vskip 0.2in 
\begin{tabular}{ lr | lr } 
\multicolumn4c{\bf Bank Balance Sheet (before)} \\
\toprule
\multicolumn2c{Assets (thousands \$)} & \multicolumn2c{Liabilities (thousands \$)} \\
\toprule
loans & 9,000,000 & deposits & 7,999,999 \\
treasuries  & 3,000,000 & savings  & 2,000,000 \\
reserves  & 149,998 & bonds  & 1,000,000 \\
cash  & 200,000 & & \\
bitcoin & 400,007 &  \\
CBDC & 149,994 & net worth & 1,900,000 \\
\bottomrule
\end{tabular}
\quad
\begin{tabular}{ lr | lr } 
\multicolumn4c{\bf Bank Balance Sheet (after)} \\
\toprule
\multicolumn2c{Assets (thousands \$)} & \multicolumn2c{Liabilities (thousands \$)} \\
\toprule
loans & 9,000,000 & deposits & {\bf 7,999,994} \\
treasuries  & 3,000,000 & savings  & 2,000,000 \\
reserves  & {\bf 149,993} & bonds  & 1,000,000 \\
cash  & 200,000 & & \\
bitcoin & 400,007 &  \\
CBDC & {\bf 149,994} & net worth & 1,900,000 \\
 \bottomrule
\end{tabular}
\vskip 0.2in 
\quad
\begin{tabular}{ lr | lr } 
\multicolumn4c{\bf Narrow Bank Balance Sheet (before)} \\
\toprule
\multicolumn2c{Assets (thousands \$)} & \multicolumn2c{Liabilities (thousands \$)} \\
\toprule
loans & 2,500,000 & savings  & 2,500,000 \\
reserves  & 11,000,002 & deposits & 10,000,001 \\
cash  & 100,000 & FBSC  & 1,000,001 \\
CBDC & 150,000 & & \\
 & & net worth & 250,000 \\
 \bottomrule
\end{tabular}
\begin{tabular}{ lr | lr } 
\multicolumn4c{\bf Narrow Bank Balance Sheet (after)} \\
\toprule
\multicolumn2c{Assets (thousands \$)} & \multicolumn2c{Liabilities (thousands \$)} \\
\toprule
loans & 2,500,000 & savings  & 2,500,000 \\
reserves  & {\bf 11,000,007} & deposits & {\bf 10,000,006} \\
cash  & 100,000 & FBSC  & 1,000,001 \\
CBDC & 150,000 & & \\
 & & net worth & 250,000 \\
\bottomrule
\end{tabular}
\caption{Balance sheets before (left) and after (right) Bob buys \$5,000 in digital trade coins (DTC) from an Administrator.}
\label{DTC1}
\end{table}

DTCs can be viewed as an instrument of barter taken to the next level. Since a DTC is anchored on a basket of assets that have real value, its price has low volatility, similar to that of the corresponding basket. As a result, it can be used as a transaction currency, as well as a unit of account and a store of value (as much as gold or oil can). Moreover, DTCs can serve as a much-needed counterpoint for national fiat currencies since their value is only indirectly affected by central banks' activities. Viewed in this way, a DTC can be seen as a privately-issued version of a supranational currency used for international trade, along the lines of Keynes's {\em bancor} (see \cite{Schumacher1943} and \cite{Keynes1943}).

\subsection{Over-collateralized stable coins (OSC)}

As a final class of stable coins, we now consider coins that are collaterilized by what we described in Section \ref{bitcoin_section} as pure-asset coins, that is to say unstable native cryptocurrencies in some existing blockchain.  

The best-known example in this class is the Dai, a decentralized cryptocurrency issued by MakerDAO, an open-source project on the Ethereum blockchain created in 2014\footnote{See \url{https://docs.makerdao.com}.}. Any user of the MakerDAO protocol (namely anyone with an identity recognized in the system) can acquire an arbitrary amount of Dai by {\em borrowing} them from the MakerDAO system, provided they lock an amount of ETH (or other assets accepted by the system) equal to a multiple (as determined by the system) of the Dai borrowed. Once acquired in these manner, Dai can be used as any other cryptocurrency, for example to buy goods and services from other agents, or be sold in exchanges. The collateral remains locked in the vault and can only be retrieved by returning the corresponding amount of Dai (that is, repying the loan), plus a Stability Fee (also set by the system) that can also only be paid in Dai. The target price of the Dai is currently set at 1 USD, and this soft peg is achieved by a combination of mechanisms. At a fundamental level, the ability of the system to create arbitrary amounts of Dai prevents its price from increasing too far away from the peg, as new Dai can be created and immediately acquired by agents who demand them (provided they lock in the necessary collateral) whenever demand rises \footnote{Compare this with the case of the tightly controlled supply of Bitcoin, which means that whenever there is higher demand, its price increases without bounds.}. On the other hand, the existence of a larger amount of collateral prevents the price of Dai from dropping too far from the peg, as it can be returned to the system in exchange of collateral whenever the demand drops\footnote{Compare again with Bitcoin, which does not have the guarantee to be converted into anything else when demand drops, leading to arbitrary price decreases.}. However, because the collateral asset (for example ETH) can itself experience large price fluctuations, the amount of overcollateralization needs to be set high enough for this mechanism to work\footnote{As illustration, the total amount of Dai in circulation in August 2021 is approximately USD 6 Billion, whereas the total amount of assets locked in MakerDAO vaults is close to USD 10 billion.}.  

Table \ref{OCS1} illustrates what happens when Alice, having already used \$10,000 from her savings to purchase Ethereum, decides to lock \$5,000 of which in a vault managed by the MakerDAO system (which continues to be owned by Alice for accounting purposes) in exchange for \$3,000 in Dai. Observe that the newly acquired Dai match an increase in liabilities for Alice in the form of a Dai loan that needs to be repaid to the system. Accordingly, the assets of MakerDAO increase by the amount of this loan, with a corresponding increase in liability in the form of Dai in circulation. Notice the close resemblance to the familiar endogenous money creation that characterizes a regular bank loan. The difference here is that the creation of new Dai is entirely decentralized and demand-driven, whereas in the case of regular bank loans, the creation of deposits, in principle, needs to be approved by the bank. Observe, however, that in reality a commercial bank will almost always extend loans to a client with sufficient collateral, provided the bank has enough capital to satisfy regulatory constraints. Because MakerDAO does not face this type of capital requirement, its ability to create Dai is even more unrestricted than in traditional endogenous money creation frameworks for banks as described, for example, in \cite{Lavoie2003}.

\begin{figure}[h!]
\centering
   \includegraphics[trim=0 7cm 0 7cm,width=\textwidth]{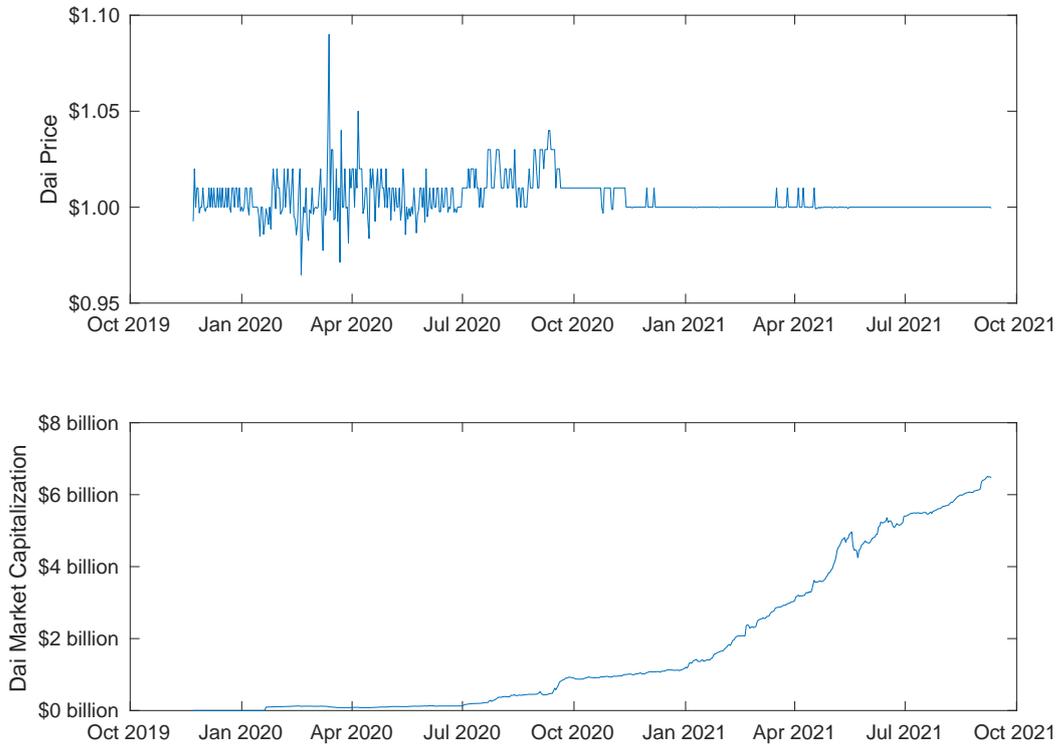}
 \caption{Price (top) and market capitalization (bottom) for Dai in USD. Source: coinmarketcap.com, downloaded on Sept 11, 2021.}
 \label{Dai_figure}
\end{figure}

Figure \ref{Dai_figure} show the price and market capitalization of Dai over the last two years. We observe that, apart from brief periods of volatility at the outset of the COVID-19 financial turbulence in 2020, the stabilization mechanism described above seems to have worked well, with the price of Dai remaining within 10\% of parity with the USD, which partially explains the rapid increase in market capitalization observed in the last year.   

\begin{table}
\centering
\begin{tabular}{ lr | lr } 
\multicolumn4c{\bf Alice's Balance Sheet (before)} \\
\toprule
\multicolumn2c{Assets (\$)} & \multicolumn2c{Liabilities (\$)} \\
\toprule
 house & 750,000 & mortgage & 500,000 \\
 car & 30,000 & car loan & 20,000 \\
 bank deposit & 500 & credit card & 1,500 \\
 savings & 18,000 & dai loan & 0 \\
bitcoin & 3,000 &  & \\
artwork & 8,000 & & \\
FBSC & 1,000 & & \\
ethereum & 10,000 & & \\
vault & 0 & & \\
dai & 0 & net worth & 299,000 \\
\bottomrule
\end{tabular}
\quad
\begin{tabular}{ lr | lr } 
\multicolumn4c{\bf Alice's Balance Sheet (after)} \\
\toprule
\multicolumn2c{Assets (\$)} & \multicolumn2c{Liabilities (\$)} \\
\toprule
 house & 750,000 & mortgage & 500,000 \\
 car & 30,000 & car loan & 20,000 \\
 bank deposit & 500 & credit card & 1,500 \\
 savings & 18,000 & dai loan & {\bf 3,000} \\
bitcoin & 3,000 &  & \\
artwork & 8,000 & & \\
FBSC & 1,000 & & \\
ethereum & {\bf 5,000} & & \\ 
vault & {\bf 5,000} \\
dai & {\bf 3,000} & net worth & 299,000 \\
\bottomrule
\end{tabular}
\vskip 0.2in 
\begin{tabular}{ lr | lr } 
\multicolumn4c{\bf MakerDAO Balance Sheet (before)} \\
\toprule
\multicolumn2c{Assets (thousands \$)} & \multicolumn2c{Liabilities (thousands \$)} \\
\toprule
dai loans & 6,000,000 & dai & 6,000,000 \\
other assets & 2,000,000 & net worth & 2,000,000 \\
\bottomrule
\end{tabular}
\quad
\begin{tabular}{ lr | lr } 
\multicolumn4c{\bf MakerDAO Balance Sheet (after)} \\
\toprule
\multicolumn2c{Assets (thousands \$)} & \multicolumn2c{Liabilities (thousands \$)} \\
\toprule
dai loans & {\bf 6,000,003} & dai & {\bf 6,000,003} \\
other assets & 2,000,000 & net worth & 2,000,000 \\
 \bottomrule
\end{tabular}
\caption{Balance sheets before (left) and after (right) Alice acquires \$3,000 in Dai by locking in \$5,000 in ETH in her vault.}
\label{OCS1}
\end{table}

\section{Towards a macroeconomic model with cryptocurrencies}
\label{SFC}

In the previous sections, we described the economic properties of different types of cryptocurrencies by means of representative examples of balance sheet transactions between agents. Naturally, these transactions correspond to only a small sample of the transactions executed by such agents, and the balance sheets themselves are incomplete snapshots of the relationships between them. For example, in Table \ref{bitcoin2}, Alice and Bob each have mortgages as their liabilities, but we do not indicate in whose balance sheets these mortgages are listed as assets, likely a different bank for each. Similarly, in Table \ref{bitcoin3}, Bob's deposits are a tiny fraction of his bank's total deposits. As yet another example, the central bank in Table \ref{CBDC1} has liabilities in the form of reserves held as assets by a large number of banks, including the Alice's narrow bank and Bob's commercial bank.

The systematic aggregation of balance sheets for all agents of a given group or sector, the changes in balance sheet arising from transactions between entire sectors, and the exploration of their macroeconomic consequences are the principles of the approach known as Stock-Flow Consistent (SFC) Modeling \cite{Santos2005}. Whereas a full specification of an SFC model incorporating cryptocurrencies is beyond the scope of this paper and will be explored elsewhere, we nevertheless find instructive to provide its basic structure, as a natural extension of the examples and definitions introduced above. 

Accordingly, consider at a minimum a seven-sector open economy consisting of Households, Firms, Banks, Other Financial Institutions, the Treasury, the Central Bank, and The Rest of the World with the following basic properties.   

\begin{enumerate}
    \item {\bf Households:} Households hold conventional cash, government debt, bank deposits, shares of firms, banks, and other financial institutions, and all types of cryptocurrencies discussed in the paper (pure-assets, CBDC, and stable coins) as assets. Their liabilities consist of bank loans, and loans provided by other financial institutions. They receive income in the form of wages, interest on deposits and government debt, and dividends, which they then use to pay taxes and consume goods and services. The difference between income received and expenses paid constitute savings, which are allocated among balance sheet items (for example increasing deposits, or paying down consumer debt).   
    
    \item {\bf Firms:} Firms hold all types of assets held by households, except shares\footnote{This is done for simplicity only. In reality firms own shares of other firms, although this would be netted at a sector level, but can also own shares of banks and other lending institutions, in the same way that banks can own shares of firms. Unless strictly necessary to highlight a specific financial phenomenon (for example fire sales of securities by banks in periods of distress) we prefer the simplifying assumption in which shares are exclusively owned by households.} in addition to physical capital\footnote{For the purpose of SFC models, following a convention adopted by national accounts in most countries, real estate is considered part of the capital in the Firms sector, with homeowners treated as sole-proprietors receiving income either as actual rent from tenants or saved rent not paid to other landlords. Accordingly, mortgage debt can be combined with Firms debt. See \url{https://www.bea.gov/sites/default/files/methodologies/nipa_primer.pdf} for details.} and inventories. Their liabilities consist of bank loans, loans provided by other financial institutions, shares\footnote{We favour the accounting convention of treating shares issued by companies as explicit liabilities, so as to match the assets held by shareholders. For details, see \cite{GodleyLavoie2007}.}, and certain types of privately issued cryptocurrencies (for example DTC in the case of the administrator of a consortium). They choose the level of production based on long-term growth and short-term fluctuations in demand, as well as the level of investment (which is itself part of demand) and receive revenue from the sales of goods and services as income (including revenue from services associated with the cryptocurrencies they issue), which they use to pay wages, depreciation costs, interest on debt, taxes, and dividends. The difference between income received and expenses paid constitute retained profits, which are allocated among balance sheet items (say an increase in capital, or loan repayment). 
    
    \item {\bf Banks:} Banks, including both commercial and narrow banks, hold loans, government debt, reserves, and potentially all types of cryptocurrencies as assets. Their liabilities consist of deposits (both sight and time deposits), shares, and certain types of privately issued cryptocurrencies (for example FBSC issued by a narrow bank). They receive fees and interest on loans, reserves, and government debt as income, with which they pay taxes, dividends, and interest on deposits. The difference between income received and expenses paid constitute retained profits, which are allocated among balance sheet items (say an increase in holding of government debt). They can create money (i.e deposits) endogenously in the case of fractional banking, but are not allowed to do so in the case of narrow banking. They interact with the central bank through purchase and sale of government debt in exchange of reserves, which in turn is how the central bank implements monetary policy.      
    
    \item {\bf Other Financial Institutions:} This sector includes the type of lending institutions that do not have demand deposits as their liabilities (for example they issue shares in order to fund their lending) that arise in the context of narrow banking, as well as providers of consumer loans and credit card debt. It also includes private issuers of stable coins, such as custodial stable coins (e.g. Tether) and overcollateralized stable coins (e.g. Dai). Their assets are loans and other debt securities, and their liabilities are shares and the cryptocurrencies they issue.  
    
    \item {\bf Treasury:} This is the part of a sovereign government that sets spending and taxation, which in the simplest form can be given by constant ratios, so that the government debt remains stable, since the main focus of such model are the operations of the central bank and interactions with cryptocurrencies. Also for simplicity, the only asset of this sector is a deposit account with the central bank, which is used for spending and transfers to other sectors, as well as to collect taxes, whereas the only liability are treasury bills (or other government debt of longer maturity), on which it pays interest. In addition, we assume that Treasury owns the Central Bank (although they can be operationally independent), and therefore receives all of its profits.  
    
    \item {\bf Central Bank:} This is a key sector in the model, whose liabilities consist of cash, reserves, the Treasury deposit account, and CBDC. Its assets consists primarily of government debt, but could also include other securities, as in the case of quantitative easing. As mentioned above, any profits for this sector (say a difference between interest earned and interest paid) are transferred to the Treasury. The Central Bank in this model performs the following functions: (i) purchase and sale of government bonds from the banking sector in order to achieve a desired level for the policy rate; (ii) issuing of CBDC; (iii) direct lending to firms and households in periods when bank credit is not sufficient. 
    
    \item {\bf Rest of the World:} In order to investigate the potential of DTC as a supranational currency for the purpose of trade, we need to include an external sector in the economy. We assume for simplicity that, from the point of view of the domestic economy, the Rest of the World can be represented
    entirely by real imports, consisting of household consumption of foreign goods and services purchased
directly from abroad, and real exports, consisting of domestic goods and services sold by firms to foreign consumers. In addition, we assume that the only domestic asset held by foreigners consist of cash,
which is a liability of the central bank, and DTC, which is a liability of the administrator that issues them on behalf of a consortium of domestic sponsors. More general versions of the model could include Treasury bills and bank deposits as assets held by foreigners, as well as many types of foreign assets held by domestic households.
\end{enumerate}

Apart from the basic accounting properties described above, a fully specified stock-flow consistent model needs to include all of the structural functions determining the behavior of each sector, such consumption of households, investment by firms, lending by banks, spending and taxation by the government, etc. We plan to pursue this line of research in a subsequent publication, and hope that the basic structure described above motivates other to do the same. 

\section{Conclusion} 

As we have seen in the previous sections, cryptocurrencies can perform multiple economic functions. Pure-asset coins such as Bitcon and Ethereum are economically more similar to precious metals and commodities, which are held primarily for hedging and speculative purposes. Others such as Tether and Dai are functionally closer to conventional bank deposits, thereby relying on the financial health of the institution issuing them for liquidity and stability, whereas CBDC are the closest type of cryptocurrency to conventional cash. 

The successes and merits of each cryptocurrency depend as much on its technological implementation as on its economic fundamentals. For example, the time and energy-consuming consensus mechanism by PoW used by Bitcoin explains its low number of transactions per second and corresponding unsuitability as a scalable medium of exchange for retail use. But as we have seen, the much simpler and faster Ripple protocol also exhibits extreme price volatility for its native token XRP, as do all pure-asset coins. The reason for this is that, by not being a liability for any economic agent and also not having any intrinsic value, these coins lack any credible mechanism for price stabilization and have to rely {\em exclusively} on the expectation that other users will want to buy or hold them at some point in the future. 

At the other extreme, CBDC are fully backed by central banks and therefore have exactly the same stability properties as fiat currencies. On the other hand, widespread adoption of digital tokens directly managed by a central bank and available to the public at large raises its own operational and economic problems, chief among them concerns regarding KYC and AML functions, as well as privacy. The approached reviewed above utilizes a partnership between the central bank and the private sector, which is tasked with administering the CBDC owned by members of the public, while these coins remain entirely a financial liability of the central bank. Implemented in this way, CBDC would have all the properties associated with cash---except anonymity with respect to the specific private sector intermediary chosen to administer one's digital wallet---plus all the conveniences associated with digital alternatives. 

Between the two extremes of pure-asset coins and CBDC, one finds the growing class of stable coins, with varying degrees of convenience and price stability. As we have seen, both FBSC (viewed by some as synthetic CBDC) and CSCs such as Tether necessitate the introduction of a narrow bank as either the issuer or the designated deposit holder for the issuer of these coins. That is to say, the old notion of full-reserve banking (see \cite{Pennacchi2012} for an overview) finds a new {\em raison d'être} in the context of cryptocurrencies. Similarly, technological innovations allowing the creation of DTC could represent the arrival of an asset-backed supranational stable currency long dreamed of by economists (see references in \cite[Section 2]{LiptonHardjonoPentland2018}). Finally, as we have seen, issuance of overcollaterilized stable coins such as Dai share all the similarity with the type of endogenous money creation in the form of bank deposits arising from loans, without the corresponding regulatory framework, at least presently. 

In conclusion, when viewed as the next chapter of a long and turbulent history, cryptocurrencies point to a future of money that is in all respects quite similar to its past: uncertain, diverse, hierarchical, hybrid, fundamental to society, and ultimately fascinating.

\bibliographystyle{siam}
\bibliography{finance}

\end{document}